\documentclass[11pt,twoside,preprintnumbers]{article}
\usepackage{axodraw}
\usepackage{pstricks}
\usepackage{color}
\usepackage{cite}
\usepackage{times}

\usepackage[T1]{fontenc}
\usepackage{amsmath}
\usepackage{dsfont}
\usepackage[latin1]{inputenc}
\usepackage{a4wide}
\usepackage{graphicx}
\usepackage{amssymb,multirow}
\usepackage{psfrag}
\usepackage{epsfig}
\usepackage{scalefnt}
\usepackage{units}
\usepackage{scrpage}  
\usepackage{fancyhdr}

\newcommand{\eq}[1]{Eq.~(\ref{#1})}
\newcommand{\eqsand}[2]{Eqs.~(\ref{#1}) and (\ref{#2})}

\newcommand{\sgn}{\mbox{sgn}\,}

\newcommand{\lt}{\left}
\newcommand{\rt}{\right}
\newcommand{\gev}{\,\mbox{GeV}}

\newcommand{\fig}[1]{Fig.~\ref{#1}}
\newcommand{\tab}[1]{Tab.~\ref{#1}}

\mathcode`,="013B
\mathcode`.="613A

\medskipamount 


\title{{\normalsize TTP12-006 \hfill February 2012}
  \\[-12pt]
  {\normalsize TUM-HEP-827/12 \hfill \phantom{space}}
  \\[12pt]
 $\Gamma\left(K\to
                  e\nu\right)/\Gamma\left(K\to\mu\nu\right)$ in the\\
                  Minimal Supersymmetric Standard Model}
\author{Jennifer Girrbach${}^{1,2,3}$ and Ulrich Nierste${}^1$ 
\\[6pt]
{\normalsize
   {\slshape $^1$ \parbox[t]{0.8\textwidth}{Institut f\"ur Theoretische
      Teilchenphysik, Karlsruhe Institute of Technology,\\
      Universit\"at Karlsruhe, Engesserstra\ss e 7, 76128 Karlsruhe, Germany}
    }}
  \\
  {\normalsize {\slshape $^2$ \parbox[t]{0.8\textwidth}{Institute of Advanced Study, Technische Universit\"at M\"unchen,\\ Lichtenbergerstra\ss{}e 2a, 85748 Garching, Germany}}}
  \\
  {\normalsize {\slshape $^3$  \parbox[t]{0.8\textwidth}{Excellence Cluster Universe, Technische Universit\"at M\"unchen,\\
Boltzmannstra\ss{}e 2, 85748 Garching, Germany  }}}
}

\date{}
\begin{document}

\maketitle

\begin{abstract}
  The quantity $R_{K} = \Gamma(K\rightarrow e \nu)/\Gamma(K\rightarrow
  \mu \nu)$ studied by the experiment NA62 at CERN is known to probe
  lepton-flavour violating (LFV) parameters of the Minimal
  Supersymmetric Standard Model (MSSM).  A non-zero parameter
  $\delta_{RR}^{13}$ can open the decay channel $K\to e \nu_\tau$ and
  enhance $R_K$ over its Standard-Model value.  In the region of the
    parameter space probed by NA62 the contribution from a bino-stau
    loop diagram is numerically dominant and the mixing between
    left-handed and right-handed staus is important. For large values of
    the stau mixing angle $\theta_\tau$ the commonly adopted mass
    insertion approximation is not accurate.  We therefore express
  the supersymmetric contribution to $R_K$ in terms of the mass of the
  lightest stau eigenstate, the mixing angle $\theta_\tau$, and other
  relevant MSSM parameter such as $\tan\beta$ and the charged-Higgs
  boson mass $M_H$ and plot the parameter regions constrained by
  $R_K$. We further study to which extent $R_K$ can be depleted through
  MSSM contributions interfering destructively with the SM amplitude for
  $K\to e \nu_e$. This lepton-flavour conserving (LFC) mechanism
  involves the parameter combination
  $|\delta_{LL}^{13}\delta_{RR}^{13}|$, which can be constrained with a
  naturalness consideration for the electron mass or with the
  measurement of the anomalous magnetic moment of the electron. The LFC
  effect on $R_K$ is marginal, an NA62 measurement of $R_K$
  significantly below the Standard Model expectation would indicate
  physics beyond the MSSM.
\end{abstract}

\section{Introduction}
The discovery of neutrino oscillations has shown that individual lepton
numbers are not conserved. This phenomenon constitutes physics beyond the
Standard Model in its original formulation, which involves only renormalisable
interactions and contains no right-handed neutrino fields. Nevertheless, the
Standard Model can accommodate neutrino oscillations with the help of a
dimension-5 term, which leads to a Majorana mass matrix for the neutrinos.  By
diagonalising this matrix one obtains the physical neutrino masses and the
Pontecorvo-Maki-Nakagawa-Sakata (PMNS) matrix encoding the strength of the
flavour transitions \cite{Pontecorvo:1957qd,Maki:1962mu}. The dimension-5 mass
term is naturally generated if right-handed neutrino fields are added to the
Standard-Model (SM) Lagrangian: being gauge singlets these fields permit
fundamental Majorana mass terms, which are not protected by the SM gauge
symmetry and can consequently be very large.  Integrating out the heavy
right-handed neutrinos generates the dimension-5 term and the desired small
neutrino masses $m_{\nu_i}$ through the famous see-saw formula
\cite{Minkowski:1977sc,Yanagida:1979as,Glashow:1979nm,GellMann:1980vs,
Mohapatra:1979ia}. With this set-up lepton-flavour violating (LFV) decays
of charged leptons like $\ell_{j}\rightarrow \ell_{i}\gamma$ (where
$\ell_{1,2,3}=e,\mu,\tau$) occur at unobservably small rates, because the
transition amplitudes are suppressed by a factor of
$(m_{\nu_j}^2-m_{\nu_i}^2)/M_W^2$.  This situation is dramatically different
in the Minimal Supersymmetric Standard Model (MSSM), which contains new
sources of flavour violation in the soft supersymmetry-breaking sector.  To
study LFV effects one commonly adopts a weak basis of the (s)lepton multiplets
in which the lepton Yukawa couplings are flavour-diagonal. The off-diagonal
elements of the charged slepton mass matrix, $\Delta m_{XY}^{ij}$ with $i,j =
1,2,3$ and $X,Y = L,R$, give rise to LFV decays of charged leptons through
loops containing a slepton and a neutralino. By confronting MSSM predictions
with experimental upper bounds on LFV decay rates one can derive constraints
on these elements, which are usually quoted for the dimensionless parameters
\begin{equation}
 \delta_{XY}^{ij} = 
 \frac{\Delta m_{XY}^{ij}}{\sqrt{m_{i_{X}}^{2}m_{j_{Y}}^{2}}} \label{defdxy}.
\end{equation}
Here $m_{i_{X}}$ denotes the $i$-th diagonal element of the slepton mass
matrix with chirality $X=L,R$. If all $\delta_{XY}^{ij}$ are small,
$m_{i_{X}}$ essentially coincides with the corresponding physical $i$-th
generation charged slepton mass.  A different avenue to constrain the
$\delta_{XY}^{ij}$ in \eq{defdxy} are studies of deviations from lepton flavour
universality (LFU). This approach has been proposed in Ref.~\cite{Masiero},
which exploits the impressive experimental precision of
\begin{equation}
 R_K =  \frac{\Gamma(K\rightarrow e \nu)}{\Gamma(K\rightarrow \mu \nu)}.
 \label{defrk}
\end{equation}
This notation implies a sum over all three neutrino species. The experimental
situation is summarised in \tab{RKexpWerte}. The cancellation of the hadronic
uncertainties makes the theoretical prediction of $R_K$ very clean: 
Including bremsstrahlung the SM value is given by
\cite{Finkemeier:1995gi,Cirigliano:2007xi,Cirigliano:2007ga}
\begin{equation}\label{equ:RKSMlit} 
R_{K}^{\rm SM} = \frac{m_{e}^{2}}{m_{\mu}^{2}} 
   \frac{(m_{K}^{2}-m_{e}^{2})^{2}}{(m_{K}^{2}-m_{\mu}^{2})^{2}}
   \left(1+\delta R_\text{QED}\right) = (2.477\pm0.001)\cdot 10^{-5}
\end{equation}
The large helicity suppression of the SM contribution to the electronic decay
mode makes $R_K$ sensitive to effects of a charged Higgs boson. In the MSSM
the charged-Higgs contribution cancels from $R_K$ at tree-level. Yet, as
pointed out in Ref.~\cite{Masiero}, at the loop level LFU-violating
contributions involving $\Delta m_{XY}^{ij}$ can lead to $R_K\neq R_K^{\rm
  SM}$. It is convenient to 
parametrise the $\mu-e$ non-universality in $R_K$ in terms of the
quantity  $\Delta r^{\mu-e}$ defined as 
\begin{equation}
 R_K = R_K^\text{SM}\left(1 + \Delta r^{\mu-e}\right) . \label{defr}
\end{equation}
Supersymmetric contributions which are linear in $\delta_{XY}^{ij}$
cannot interfere with the SM amplitude in $K\to \ell \nu_\ell$,
because they lead to a final state with charged lepton and neutrino
belonging to different fermion generations. Therefore these
contributions will necessarily increase $\Gamma(K\to \ell \nu)$. In
Ref.~\cite{Masiero} a mechanism involving the product
$\delta_{LL}^{13}\delta_{RR}^{13}$ has been proposed to achieve a
suppression of $\Gamma(K\to e \nu_e)$ and therefore of
$R_K$. Recently, two new observables have been found to constrain the
very same combination of supersymmetric FCNC parameters
\cite{Girrbach:2009uy}: Firstly, 't Hooft's naturalness criterion has
been applied to the electron mass yielding a non-decoupling upper
bound on $|\delta_{LL}^{13}\delta_{RR}^{13}|$. Secondly, a powerful
  bound on $|\delta_{LL}^{13}\delta_{RR}^{13}|$  
has been derived from the anomalous magnetic moment of the
electron.  The latter constraint decouples, i.e.\ becomes weaker for
larger superpartner masses, but the bounds on
  $|\delta_{LL}^{13}\delta_{RR}^{13}|$ are comparable to the ones found
  from the electron mass for typical sparticle spectra.
In Sect.~\ref{sect:lfc} we use the results
of Ref.~\cite{Girrbach:2009uy} to assess the possible maximal effect
on $R_K$ from loop diagrams involving
$\delta_{LL}^{13}\delta_{RR}^{13}$. In Sect.~\ref{sec:LFVself-energy} 
we study MSSM contributions involving a single power of 
either $\delta_{LL}^{13}$ or $\delta_{RR}^{13}$. These contributions 
feed $K\to e \nu_\tau$ and therefore increase $R_K$. In
Sect.~\ref{sect:c} we conclude. 
\begin{table}[t]
\begin{center}
 \begin{tabular}{|c|c|c|}
 \hline
  Experiment & $R_{K}$ [$10^{-5}$] & error $\nicefrac{\delta R_K}{R_K}$\\
 \hline
 \hline
 PDG 2006 & $2.45 \pm 0.11$& 4.5\%\\
\hline
 NA48/2 2003  &$2.416\pm0.043\pm0.024$&2.8\%\\
 NA48/2 2004  & $2.455\pm0.045\pm0.041$& 3.5\%\\
 KLOE  & $2.55\pm0.05\pm0.05$& 3.9\%\\
 \hline
 Kaon 2007 & $2.457\pm0.032$&1.3\%\\
 \hline
 PDG 2008 & $2.447\pm 0.109$ & 4.5\%\\
PDG 2010 & $2.493\pm 0.036$ & 1.4\%\\
\hline
KLOE 2009 & $2.493\pm0.025\pm0.019$&1.7\%\\
NA62 Jan 2011 & $2.487\pm0.013$& 0.5\%\\
NA62 Jul 2011 & $2.488\pm0.010$& 0.4\%\\
\hline\hline 
 \end{tabular}
\end{center}
\caption{Experimental values for $R_{K}$
  \cite{PDG,Isidori:2007zs,Collaboration:2009rv,Goudzovski:2010uk, 
        Goudzovski:2009talk,:2011uv,Goudzovski:2011talk,Goudzovski:2011tn,
        Goudzovski:2011tc}.
We use the published result of Ref.~\cite{:2011uv} quoted in the
  second-to-last row, which corresponds to $\Delta r^{\mu-e}=0.004\pm
  0.005$. The result in the last row, reported by NA62 at
  conferences \cite{Goudzovski:2011tn,Goudzovski:2011tc}, corresponds 
  to the full data set collected by NA62 in 2007--2008.
For the future an experimental accuracy of $\delta R_K/R_K =0.1-0.2\%$
is feasible for the NA62 experiment
\cite{Goudzovski:2009talk}.}\label{RKexpWerte}~\\[-2mm]\hrule
\end{table}

\section{Lepton-flavour conserving loop corrections\label{sect:lfc}}
In a two-Higgs-doublet model of type II the following Hamiltonian
describes leptonic Kaon decays \cite{Hou:1992sy}:
\begin{equation}
\mathcal{H}= \frac{G_{F}}{\sqrt{2}}V_{us}
  \left[\overline{u}\gamma^{\mu}(1-\gamma^{5})s
  \overline{\nu_\ell}\gamma_{\mu}(1-\gamma^{5})\ell-\frac{m_{s}m_\ell}{
  M_{H}^{2}}\tan^{2}\beta\,\overline{u}(1+\gamma^{5})s
  \overline{\nu_\ell}(1-\gamma^{5})\ell\right] \label{defh}
\end{equation}
yielding the decay rate 
\begin{equation}        
\Gamma(K\rightarrow \ell\nu_\ell) =
      \frac{G_{F}^2}{8\pi}m_\ell^{2}m_{K}f_{k}^{2}|V_{us}|^{2}\left(
      1-\frac{m_\ell^{2}}{m_{K}^{2}}\right)^{2}\left[
      1-m_{K}^{2}\frac{\tan^{2}\beta}{M_{H}^{2}} \right]^{2},
      \qquad \ell = e,\,\mu \label{decay}
\end{equation}
where the second contribution in the square brackets stems from the
additional charged-Higgs exchange.  At tree level the relative Higgs
contribution to the decay rate is independent of the lepton flavour
and thus cancels in the ratio $R_{K}$ defined in \eq{defrk}.  However,
SUSY loop corrections can introduce a dependence on the lepton
flavour: In the large $\tan\beta$ regime of the MSSM the relation
between the Yukawa couplings and the measured fermion masses can
change significantly, with the loop suppression compensated by a
factor of $\tan\beta\sim 50$. In the decoupling limit $M_\text{SUSY}\gg
v,M_H$ these enhanced corrections arise in a very intuitive way from a
loop-induced non-holomorphic Higgs coupling \cite{Hall:1993gn}. (Here
$\tan \beta=v_u/v_d$ denotes the ratio of the two Higgs vevs,
$v=\sqrt{v_u^2+v_d^2}=174\gev$ is the electroweak scale, $M_H$
represents the charged-Higgs boson mass, and $M_\text{SUSY}$ is the mass
scale of the supersymmetric particles entering the loop diagrams.) In
our case of lepton Yukawa couplings the applicability of the
decoupling limit is not clear a priori, because some of the
superpartners involved (e.g.\ neutralinos) can easily have
masses around or even below $v$. To cover the case $M_\text{SUSY}\sim
v,M_H$ one must resort to a diagrammatic resummation of $\tan
\beta$-enhanced corrections, which has been worked out for quarks in
Refs.~ \cite{Carena:1999py,Hofer:2009xb} and for leptons in
Refs.~\cite{Marchetti:2008hw,Girrbach:2009uy}. The desired all-order
relation between the Yukawa coupling $y_\ell$ and the physical lepton
mass $m_\ell$ is
\begin{equation}
  -y_\ell v_d = m_\ell + \Sigma_\ell - \Sigma_\ell^2 + \Sigma_\ell^3 - \ldots  
 \label{geo}
\end{equation}
where $\Sigma_\ell$ is the piece of the one-loop self-energy proportional
to $m_\ell\tan \beta$.  In the on-shell renormalisation scheme the mass
counterterm is just $\delta m_\ell = \Sigma_\ell$ up to terms which are
not enhanced by a factor of $\tan \beta$. \eq{geo} is conventionally
written as 
\begin{equation}
  -y_\ell \sin \beta \; =\;
      \frac{m_\ell\tan\beta }{v(1+\epsilon_{\ell}\tan\beta)}
      \;=\;  \frac{g_{2}}{\sqrt{2}M_{W}} \,
        \frac{m_\ell\tan\beta }{1+\epsilon_{\ell}\tan\beta}, \qquad\qquad
  \epsilon_{\ell}\tan\beta = -\frac{\Sigma_{\ell}}{m_{\ell}} .\label{ysum}
\end{equation}
$-y_\ell \sin \beta$ is just the 
Higgs coupling to right-handed down-type leptons contributing to the
second terms in the square brackets in \eqsand{defh}{decay}. 
Putting everything together one finds
\begin{equation}
 R_{K}= R_{K}^\text{SM}\left[\frac{ 
     1-m_{K}^{2}\dfrac{\tan^{2}\beta}{M_{H}^{2}}
 \dfrac{1}{(1+\epsilon_{s}\tan\beta)(1+\epsilon_{e}\tan\beta)} }{
 1-m_{K}^{2}\dfrac{\tan^{2}\beta}{M_{H}^{2}}
 \dfrac{1}{(1+\epsilon_{s}\tan\beta )(1+\epsilon_{\mu}\tan\beta )}
 }\right]^{2},
\end{equation}
where $\epsilon_s$ is the analogue of $\epsilon_{\ell}$ for the strange 
Yukawa coupling. This reads
\begin{equation}
 \Delta r^{\mu-e} =
  - \frac{2m_{K}^{2}\tan^{2}\beta}{M_{H}^{2}|1+\epsilon_{s}\tan\beta|}
     \left[\frac{1}{
     |1+\epsilon_e\tan\beta|}-\frac{1}{|1+\epsilon_\mu\tan\beta|}\right]
\end{equation}
in terms of the notation of \eq{defr}.  Lepton universality is
violated for $\epsilon_e \neq \epsilon_\mu $. 
In the MSSM with minimal flavour violation (MFV) the only source 
of $\epsilon_e \neq \epsilon_\mu $ are different values of the 
selectron and smuon masses. A sizable slepton mass splitting between
the first and second generation is theoretically hard to justify and 
we do not consider this possibility any further. 

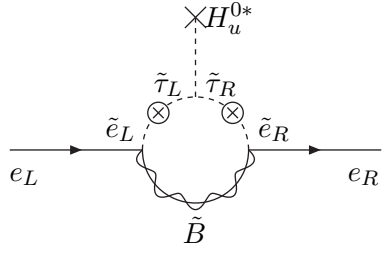
\begin{figure}[t]
  \begin{center}
  \scalebox{1}{%
    \begin{picture}(130,100)(30,10)
      \ArrowLine(30,50)(80,50) \ArrowLine(120,50)(170,50)
      \Text(30,40)[l]{$e_{L}$} \Text(170,40)[r]{$e_{R}$}
      \DashLine(100,100)(100,70){2}
      \Line(96,96)(104,104) \Line(96,104)(104,96)
      \Text(113,100)[]{$H_{u}^{0*}$}
      \DashCArc(100,50)(20,0,180){2}
      \Text(90,75)[]{$\tilde{\tau}_{L}$}
      \Text(110,75)[]{$\tilde{\tau}_{R}$}
      \BCirc(114,64){4}\BCirc(86,64){4}
      \Line(112,62)(116,66)\Line(112,66)(116,62)
      \Line(84,62)(88,66)\Line(84,66)(88,62) 
      \CArc(100,50)(20,180,0)
      \Text(72,58)[]{$\tilde{e}_{L}$}
      \Text(130,58)[]{$\tilde{e}_{R}$}
      \PhotonArc(100,50)(20,180,0){2.5}{4.5}
      \Text(100,20)[]{$\tilde{B}$}
      \Text(30,100)[l]{}
    \end{picture}
  }%
\end{center}
\caption{Dominant double LFV contribution to the electron mass 
renormalisation.}\label{doppelteLVFdominant}~\\[-2mm]\hrule
\end{figure}

An a priori sizable source of lepton non-universality are the diagrams
involving a double insertion of LFV mass insertions \cite{Masiero},
see \fig{doppelteLVFdominant}\footnote{The SU(2) partner diagram of
Fig.~\ref{doppelteLVFdominant} involving a charged Higgs boson was
shown in Ref.~\cite{Masiero}.}.
Instead of the tree level coupling which is proportional to
$m_{e}\tan\beta$ the diagram gives a contribution proportional to
$m_{\tau}\tan\beta$.  This is the dominant contribution to the LFC
self energy through double LFV and is given as (with the notation and
conventions explained in Appendices A, B of
Ref.~\cite{Girrbach:2009uy})\footnote{Interchanging the
chiralities of the two $\tilde{\tau}$ yields the analogous expression
with $\delta_{LR}^{13}\delta_{LR}^{31} $. }:
\begin{eqnarray}\label{Sigma_{e}^{FV}}
 \Sigma_{e}^{FV} &=& \frac{\alpha_{1}}{4\pi}\mu M_{1}
 m_{\tilde{e}_{L}} m_{\tilde{e}_{R}} m_{\tilde{\tau}_{L}}
 m_{\tilde{\tau}_{R}} \delta_{LL}^{13}\delta_{RR}^{13}
 \frac{m_{\tau}\tan\beta}{1 + \epsilon_{\tau}\tan\beta}
 F_{0}(M_{1}^{2},m_{\tilde{e}_{L}}^{2},m_{\tilde{e}_{R}}^{2},
       m_{\tilde{\tau}_{L}}^{2},m_{\tilde{\tau}_{R}}^{2}).
\end{eqnarray}
Throughout this paper we choose $\mu$, the gaugino mass parameters,
the trilinear SUSY breaking terms, and all off-diagonal slepton mass
matrix elements real. Therefore $\epsilon_s$, $\epsilon_\ell$ are real
as well. \eq{Sigma_{e}^{FV}} describes a non-decoupling effect, because
$F_0\propto M_{\text{SUSY}}^{-6}$. The contribution of
$\Sigma_{e}^{FV}$ to the resummation formulae in \eq{ysum}  is
\cite{Girrbach:2009uy}:
\begin{align}
 - y_e \sin \beta & =   
       \frac{g_{2}}{\sqrt{2}M_{W}} \,
       \frac{\lt( m_e + \Sigma^\text{FV}_e\rt)\tan\beta }{
              1+\epsilon_{e}\tan\beta}. 
\end{align}
Writing 
\begin{align}
\Sigma^{FV}_{e}& \equiv
 \frac{m_{\tau} \tan\beta}{1+\epsilon_{\tau}\tan\beta}\Delta^{e}_{ LR }
\end{align}
to adopt a notation similar to Ref.~\cite{Masiero} the charged-Higgs
coupling to the electron changes to
\begin{align}
  -y_e \sin \beta & = 
  \frac{ig_{2}}{\sqrt{2}M_{W}}
    \frac{m_{e} }{1+\epsilon_{e}\tan\beta} \tan\beta 
  \left(1+ \frac{\Sigma^{FV}_{e}}{m_e} \rt) \label{enhy} \\
& = 
  \frac{ig_{2}}{\sqrt{2}M_{W}}
    \frac{m_{e} }{1+\epsilon_{e}\tan\beta} \tan\beta 
  \left(1+\frac{m_{\tau}}{m_{e}}
      \frac{\tan\beta}{1+\epsilon_{\tau}\tan\beta} \Delta^{e}_{LR} 
  \right).
\end{align}
Recalling the discussion after \eq{geo} we see that there is a
one-to-one correspondence between the electron mass counterterm
$\Sigma^{FV}_{e}$ and the enhanced Yukawa coupling in \eq{enhy}. In
Ref.~\cite{Girrbach:2009uy} 't~Hooft's naturalness criterion has been
applied to to the electron mass to derive a bound on
$|\Sigma^{FV}_{e}|$: Within theories with MFV the smallness of $m_e$ is
justified by the chiral symmetry gained in the limit $y_e\to 0$.  In our
case a second symmetry-breaking parameter, $y_\tau
\delta_{LL}^{13}\delta_{RR}^{13} $ is present, and the naturalness
principle forbids large accidental cancellations between the two
contributions to $m_e$. Demanding $|\Sigma^{FV}_{e}|\lesssim m_e$ one
finds the bounds in \tab{LFVRes} \cite{Girrbach:2009uy}, which can be
summarised as
\begin{align}
 \left|\delta_{LL}^{13}\delta_{RR}^{13}\right|\lesssim 0.1 \; .
 \label{dlldrr}
\end{align}
The authors of Ref.~\cite{Masiero} have used
$\Delta_{LR}^{e}=\mathcal{O}(10^{-4})$ to bring $R_K$ into better
agreement with the NA48/2 result of 2004, see \tab{RKexpWerte}.  This
value implies a 2000\% change in the electron mass which is
incompatible with the naturalness principle. 

After considering the MSSM we turn our argument into a model-independent
analysis: For the naturalness bound on $y_e$ it is inessential in
which theory the self-energy $\Sigma^{FV}_{e}$ is calculated. Any
theory with a tree-level Higgs sector corresponding to a type-II 
2HDM and a self-energy contribution $\Sigma^{FV}_{e}$ not proportional
to $y_e$ affects the charged-Higgs coupling through the finite 
counterterm 
\begin{align}
\delta y_e & = \frac{y_e}{m_e}\delta m_e 
= \frac{y_e}{m_e} \frac{\Sigma_e^{FV} - \epsilon_e\tan\beta}{1 + \epsilon_e \tan\beta}
\simeq \frac{y_e}{m_e} \Sigma^{FV}_{e}.
\end{align}
If we take the upper bound from the fine tuning
argument $|\delta m_e| = m_e$, the allowed range for $y_e$ lies
between 0 and twice the tree-level value $m_e/v_d$. The largest 
allowed value for $-\Delta r^{\mu-e}$ in \eq{defr} therefore
corresponds to 
\begin{equation}
 \Delta r^{\mu-e}_\text{min,LFC}= 
  -4 \frac{m_K^2\tan^2\beta}{M_H^2 (1+\epsilon_s \tan\beta)}.
 \label{demin}
\end{equation}
This bound assumes that (as in the MSSM) the muon Yukawa coupling is not
substantially affected. Taking $\tan\beta = 50$,
$\epsilon_s\tan\beta=0.3$ (which corresponds to a
  typical loop suppression factor $\epsilon_s = \frac{1}{16\pi^2}$ and
  $\tan\beta = 50$), and a charged-Higgs mass of $M_H = 300~\text{GeV}$
we find $\Delta r^{\mu-e}_\text{min,LFC}= -5\cdot 10^{-3} $, which
can be probed by NA62 but is not in the $5\sigma$ discovery reach
of this experiment. Of course our consideration equally applies
  to positive values of $\Delta r^{\mu-e}_{\text{min,LFC}}$, i.e.\ the
naturalness bound implies $|\Delta r^{\mu-e}_\text{LFC}|\leq 5\cdot
10^{-3}$. We have discussed negative contributions, because $\Delta
r^{\mu-e}<0$ is an unambigous sign of an LFC mechanism, while $\Delta
r^{\mu-e}>0$ can be more easily accomodated with LFV new physics as
analysed in Sect.~\ref{sec:LFVself-energy}.  In
  Ref.~\cite{Masiero:2008cb} (which is an update of Ref.~\cite{Masiero})
  a thorough analysis of several observables in quark and lepton flavour
  physics has been performed.  While most of the points in the scatter
  plots of that paper satisfy the constraint from \eq{demin}, an
  inclusion of \eq{dlldrr} into the analysis would eliminate the
  outliers in these plots.  Further the use of \eq{dlldrr} would make
  the results of Ref.~\cite{Masiero:2008cb} less dependent on the
  anomalous magnetic moment of the muon, whose theoretical prediction in
  the SM involves uncertainties which are not fully understood.
   
We note that $\left|\delta_{LL}^{13}\delta_{RR}^{13}\right|$ can also be
bounded in a completely different way: The anomalous magnetic moment of
the \emph{electron}\ gives essentially the same bound as \eq{dlldrr} for
$M_{\rm SUSY}=500\gev$ and involves the same supersymmetric particles in
the loop as $\Sigma^{FV}_{e}$ as shown in
  Ref.~\cite{Girrbach:2009uy}. Thus relaxing the naturalness bound
$|\Sigma^{FV}_{e}|\leq m_e$ to lower $\Delta r^{\mu-e}_\text{min,LFC}$
in \eq{demin} requires the choice of larger bino or selectron masses to
comply with the electron magnetic moment. Should future NA62 data
  point towards $\Delta r^{\mu-e}<0$, an analysis in conjunction with the
  electron magnetic moment will place correlated lower bounds on these
  sparticle masses.

While NA62 is gaining statistics in the forthcoming years, we may expect
increasingly better information on $M_H$ and $\tan\beta$ from LHC
experiments, so that the bound $\Delta r^{\mu-e}_\text{min,LFC}= -5\cdot
10^{-3} $ quoted after \eq{demin} may eventually become tighter. Any
future NA62 measurement of $\Delta r^{\mu-e}$ below $\Delta
r^{\mu-e}_\text{min,LFC}$ will then establish a more exotic new physics
explanation than type-II charged-Higgs exchange, such as t-channel
leptoquark exchange.

 \begin{table}
   \centering
 \scalebox{0.85}{%
 \begin{tabular}{|c|c||c|c|c|c|c|}
     \hline
     \multicolumn{2}{|c||}{scenario} & $x=0.3$ & $x=1$ & $x=1.5$ & $x=3.0$
     & for 
     \\
     \hline
     \multirow{2}{*}{1} & 
     \multirow{2}{*}{$M_{1} = M_{2} = m_{L} = m_{R}$} & 
     0.261 & 0.073 & 0.050 & 0.026 &
     $\delta_{RR}^{13}\delta_{LL}^{13}>0$ \\
     & & 0.234 & 0.059 & 0.040 & 0.023 & 
     $\delta_{RR}^{13}\delta_{LL}^{13}<0$ 
     \\
     \hline
     \multirow{2}{*}{2} & 
     \multirow{2}{*}{$3 M_{1} = M_{2} = m_{L} = m_{R}$} & 
     0.301 & 0.083 & 0.057 & 0.029 &
     $\delta_{RR}^{13}\delta_{LL}^{13}>0$ \\
     & & 0.269 & 0.067 & 0.045 & 0.024 &
     $\delta_{RR}^{13}\delta_{LL}^{13}<0$
     \\
     \hline
     \multirow{2}{*}{3} & 
     \multirow{2}{*}{$M_{1} = M_{2} =  3 m_{L} = m_{R}$} &
     0.292 & 0.082 & 0.057 & 0.031 &
     $\delta_{RR}^{13}\delta_{LL}^{13}>0$ \\
     & & 0.235 & 0.067 & 0.042 & 0.027 &
     $\delta_{RR}^{13}\delta_{LL}^{13}<0$
     \\
     \hline
     \multirow{2}{*}{4} & 
     \multirow{2}{*}{$M_{1} = M_{2} = \frac{m_{L}}{3} =  m_{R}$} &
     0.734 & 0.210 & 0.142 & 0.071 &
     $\delta_{RR}^{13}\delta_{LL}^{13}>0$ \\
     & & 0.702 & 0.190 & 0.127 & 0.064 &
     $\delta_{RR}^{13}\delta_{LL}^{13}<0$
     \\
     \hline
     \multirow{2}{*}{5} & 
     \multirow{2}{*}{$3 M_{1} = M_{2} = m_{L} = 3 m_{R}$}  & 
     0.731 & 0.205 & 0.137 & 0.067 & 
     $\delta_{RR}^{13}\delta_{LL}^{13}>0$ \\
     & & 0.693 & 0.179 & 0.116 & 0.054 &
     $\delta_{RR}^{13}\delta_{LL}^{13}<0$
     \\
     \hline
   \end{tabular}
 }
 \caption{Different mass scenarios and the corresponding upper bounds
    for $\left|\delta_{RR}^{13}\delta_{LL}^{13}\right|$. $m_{R,L}$
    denotes the average right and left-handed slepton mass,
    respectively, $M_{1}$ and $M_2$ the bino and wino masses and $x =
    \mu/m_R$. In all scenarios, $\tan\beta=50$ and 
   $\sgn\mu=+1$.}\label{LFVRes} ~\\[-2mm]\hrule
 \end{table}

\section{Lepton-flavour violating loop corrections}\label{sec:LFVself-energy}
Flavour-violating self-energies in the charged-lepton line can induce
the decays $K\to \ell \nu_{\ell^\prime}$ with $\ell\neq \ell^\prime$
\cite{Masiero}.  A sizable effect is only possible in $K\to e \nu_\tau$,
so that LFV self-energies can only increase $R_K$. In this section we
estimate the maximal effect of lepton-flavour violating loop corrections
to $R_K$. A large correction to $R_K$ involves large $\tilde
\tau_L$--$\tilde \tau_R$ mixing. We cannot rely on the expansion in
$v^2/M_{\rm SUSY}^2$ adopted in Ref.~\cite{Masiero,Masiero:2008cb} in
this region of the MSSM parameter space, because the $\tilde \tau$
mixing angle $\theta_\tau$ vanishes for $v/M_{\rm SUSY}\to 0$. We use
the exact formulae of Ref.~\cite{Girrbach:2009uy}, which express $\Delta
r^{\mu-e}$ in terms of the masses of the physical stau eigenstates
$\tilde\tau_{1,2}$ rather than the diagonal elements
$m^2_{\tilde\tau_{L,R}}$ of the stau mass matrix. Since NA62 runs
concurrently with the LHC, it is anyway useful to express $\Delta
r^{\mu-e}$ in terms of the physical quantities probed in high-$p_T$
physics.

\begin{figure}[t]
\begin{minipage}{0.49\textwidth}

\begin{small}
 \begin{center}
\scalebox{1}{
    \begin{picture}(150,100)(-25,0)
      \SetColor{Black}
        \ArrowLine(-25,15)(10,50)\Text(-5,22)[]{$\nu_{\tau}$}
        \DashLine(10,100)(10,50){2} \Text(0,90)[]{$H^-$}
      \ArrowLine(10,50)(50,50) \ArrowLine(100,50)(140,50)
      \Text(20,40)[]{$\tau_{R}$} \Text(40,40)[]{$\tau_{L}$}\Text(130,40)[]{$e_{R}$}
        \Line(25,45)(35,55)\Line(25,55)(35,45)
      \DashLine(40,85)(75,50){2} \Line(40,80)(40,90)
      \Line(35,85)(45,85) \Text(37,95)[]{$H_{u}^*$}
      \CCirc(75,50){25}{White}{White}
      \DashArrowArc(75,50)(25,0,180){2} \BCirc(75,75){5}
      \Line(72,72)(78,78) \Line(72,78)(78,72)
      \Text(45,67)[]{$\tilde{\tau}_{1,2}$}
      \Text(60,80)[]{$\tilde{\tau}_{1,2}$}
      \Text(100,75)[]{$\tilde{e}_{R}$}
      \CArc(75,50)(25,180,360) \PhotonArc(75,50)(25,180,360){3}{5.5}
      \Text(75,20)[]{$\tilde{B}$}

\end{picture}
}
\end{center}
\end{small}
\end{minipage}
\begin{minipage}{0.49\textwidth}
\begin{small}
 \begin{center}
\scalebox{1}{
    \begin{picture}(240,100)(140,0)
      \SetColor{Black}
        \ArrowLine(125,15)(160,50)\Text(145,22)[]{$\nu_{\tau}$}
      \DashLine(160,100)(160,50){2}\Text(150,90)[]{$H^-$}
      \ArrowLine(160,50)(200,50) \ArrowLine(250,50)(290,50)
      \Text(170,40)[]{$\tau_{R}$} \Text(280,40)[]{$e_{R}$}
        \Line(175,45)(185,55)\Line(175,55)(185,45)\Text(190,40)[]{$\tau_{L}$}
      \CCirc(225,50){25}{White}{White}
      \DashArrowArc(225,50)(25,0,180){2} \BCirc(225,75){5}
      \Line(222,72)(228,78) \Line(222,78)(228,72)
      \Text(198,75)[]{$\tilde{\tau}_{1,2}$}
      \Text(250,75)[]{$\tilde{e}_{R}$}
      \CArc(225,50)(25,180,360) \PhotonArc(225,50)(25,270,360){3}{3.5}
      \Text(200,30)[]{$\tilde{H}$} \Text(252,30)[]{$\tilde{B}$}
      \DashLine(225,25)(225,5){2} \Line(220,10)(230,0)
      \Line(220,0)(230,10) \Text(240,5)[]{$H_{u}^*$}
\end{picture}
}
\end{center}
\end{small}
\end{minipage}
\caption{Dominant LFV self energies in external legs leading to flavour
non-diagonal charged-Higgs 
couplings. $\tilde{B}$ and $\tilde{H}$ denote the relevant bino and 
higgsino components of the neutralinos $\chi^0_1\ldots \chi^0_4$, 
respectively.}\label{fig:chargedHiggsLFV}~\\[-5mm]\hrule
\end{figure}
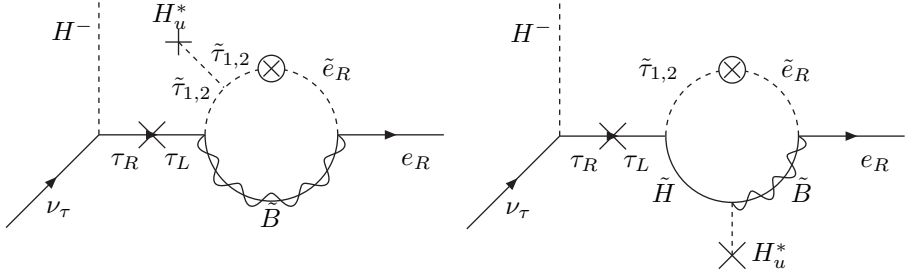

$ \Sigma_{\ell_{jR}-\ell_{iL}}$ is relevant for the $W$-coupling to
  leptons if $j>i$ (and thus contributes to threshold corrections of the
  PMNS matrix as studied in
  Ref.~\cite{Girrbach:2009uy,Crivellin:2010gw}), whereas for $j<i$ $
  \Sigma_{\ell_{jR}-\ell_{iL}}$ is responsible for the correction of the
  charged-Higgs coupling. The charged-Higgs couplings to leptons $
  \Gamma^{H^+}_{\ell_i\nu_{\ell_j}}$ including an analytic resummation
  of $\tan\beta$-enhanced corrections are listed in Eqs.~(31a-c),
  (32a-c) and (33a-c) of Ref.~\cite{Girrbach:2009uy}. For the decoupling
  limit $M_\text{SUSY}\gg v$ these charged-Higgs couplings were derived
  earlier in Eqs.~(92-95) of Ref.~\cite{Hisano:2008hn}, which further
  uses the iterative procedure of Ref.~\cite{Buras:2002vd} to resum the
  $\tan\beta$-enhanced terms.  In this paper we are interested in the
  $\tau_{L} \to e_{R}$ self-energy contributing to $R_K$ as shown in
  Fig.~\ref{fig:chargedHiggsLFV}, aiming at constraints on
  $\delta_{RR}^{31}$ to be obtained from future NA62 data.  With the LFV
  Higgs couplings at hand one can calculate the decay rates summing over
  all neutrino species and compute the ratio.  The $H^{+} e \nu_{\tau}$
  vertex, which involves the enhancement factor of $m_{\tau}/m_e$, is
  the only relevant contribution to $R_{K}$: Corrections to the muonic
  decay mode are irrelevant due to the much smaller enhancement factor
  of $m_{\tau}/m_\mu$. In view of the result of the previous section we
  can further rule out large effects in $ \Gamma^{H^{+}}_{e\nu_{e}}$. In
  the LFV case the deviation from the SM is essentially given as:
\begin{equation}
 \Delta r^{\mu-e}_\text{LFV} = \frac{m_K^4 \tan^4\beta}{M_H^4\left(1 +
     \epsilon_s\tan\beta \right)^2\left(1 +
     \epsilon_\tau\tan\beta \right)^2 }
 \frac{m_\tau^2}{m_e^2}
\left[\frac{\Sigma_{\tau_{L}-e_{R}}^{\tilde{\chi}^0}}{m_\tau }\right]^2.
\label{rme}
\end{equation}

Here $\Sigma_{\tau_{L}-e_{R}}^{\tilde{\chi}^0}$ denotes the sum 
of the two self-energies appearing in \fig{fig:chargedHiggsLFV}.
For the $\tau_L\to e_R$-transition only the two diagrams in
Fig.~\ref{fig:chargedHiggsLFV} are not suppressed with $y_e$.  We
explicitly account for stau mixing with mixing angle $\theta_\tau \in
[-\pi/4,\,\pi/4]$  and stau mass eigenstates 
$\tilde{\tau}_{1}=\tilde{\tau}_L \cos \theta_\tau 
- \tilde{\tau}_R \sin \theta_\tau$, $\tilde{\tau}_{2}=\tilde{\tau}_L 
\sin \theta_\tau  + \tilde{\tau}_R \cos \theta_\tau$. 
We use the following identities (with $m_\tau^{(0)} =
m_\tau/(1+\epsilon_\tau \tan\beta)$ and the trilinear $A$-term set to
zero since it is not $\tan\beta$ enhanced):
\begin{align}\label{equ:staumixangle}
 &\sin(2\theta_\tau) = \frac{2\Delta
    m_{LR}^{33}}{m_{\tilde{\tau}_{1}}^{2}-m_{\tilde{\tau}_{2}}^{2}}=
  \frac{-2
    m_\tau^{(0)}\mu\tan\beta}{m_{\tilde{\tau}_{1}}^{2}- 
    m_{\tilde{\tau}_{2}}^{2}},\qquad\quad
  \cos(2\theta_\tau) = \frac{m_{\tilde{\tau}_{L}}^{2}-
    m_{\tilde{\tau}_{R}}^{2}}{m_{\tilde{\tau}_1}^{2}-
    m_{\tilde{\tau}_2}^{2}}\\ &m_{\tilde{\tau}_{1,2}}^2=\frac{1}{2}\left(m_{\tilde{\tau}_{L}}^{2}+m_{\tilde{\tau}_{R}}^{
    2 } \pm\sqrt { (m_ { \tilde{\tau}_ { L } } ^
    2-m_{\tilde{\tau}_{R}}^{2})^2+4|\Delta
    m_{LR}^{33}|^2}\right)\label{equ:staueigenvalues}\\ &\text{sgn}\left(m_{\tilde{\tau}_1}^2-m_{\tilde{\tau}_2}^2\right)
  =
  \text{sgn}\left(m_{\tilde{\tau}_{L}}^{2}-m_{\tilde{\tau}_{R}}^2\right)
\end{align}
 The correlation of $\text{sgn}\left(m_{\tilde{\tau}_1}^2-m_{\tilde{\tau}_2}^2\right)$
  with $\text{sgn}\,\mu$ and $\text{sgn}\,\theta_\tau$
  can be read off from
\begin{align}
  m_{\tilde{\tau}_1} = m_{\tilde{\tau}_2} -
\frac{2\mu m_\tau^{(0)}\tan\beta}{\sin(2\theta_\tau)}\,.
\end{align}
Since we want to use the mixing angle and the mass of the lightest stau,
$m_{\tilde{\tau}_l}$, as
inputs we express the masses of the left- and right-handed stau in terms of
$m_{\tilde{\tau}_l}^2$ and $\theta_\tau$:
\begin{align}
 m_{\tilde{\tau}_L}^2 &= m_{\tilde{\tau}_l}^2 +
\frac{\mu m_\tau^{(0)}\tan\beta}{|\sin(2\theta_\tau)|}\left(1 -\text{sgn}(\theta_\tau)
\cos(2\theta_\tau)\right)\,,\\
 m_{\tilde{\tau}_R}^2 &= m_{\tilde{\tau}_l}^2 +
\frac{\mu m_\tau^{(0)}\tan\beta}{|\sin(2\theta_\tau)|}\left(1 +\text{sgn}(\theta_\tau)
\cos(2\theta_\tau)\right)\,.
\end{align}
The phenomenologically interesting large values of $\Delta r^{\mu-e}$ 
involve large values of $|\mu|$. Varying  
$|\mu|$ to larger values with $m_{\tilde{\tau}_{L,R}}^2$ fixed increases the 
mass splitting between the two stau mass eigenstates $\tilde{\tau}_l$ and 
$\tilde{\tau}_h$ and 
will eventually 
lower the smaller physical stau mass $m_{\tilde{\tau}_l}$ below its 
experimental lower bound. We avoid this problem by varying the 
parameters for fixed $m_{\tilde{\tau}_l}$. 
Treating the flavour violating off-diagonal elements
up to linear order and including stau mixing we get for the left diagram in
Fig.~\ref{fig:chargedHiggsLFV}:
\begin{align}\label{equ:tauLeRbino}\begin{split}
  \Sigma_{\tau_{L}-e_{R}}^{\tilde{B}}  =& -\frac{\alpha_{1}}{4\pi} M_1 \, m_{\tilde{e}_{R}}\, m_{\tilde{\tau}_{R}}\,\delta_{RR}^{13}\,
\sin\theta_\tau \cos\theta_\tau\left(
f_1\left(M_{1}^{2},\,m_{\tilde{e}_{R}}^{2},\,m_{\tilde{\tau}_{1}}^{2}\right)-f_1\left(M_{1
}^{2},\,m_{\tilde{e}_{R}}^{2},\,m_{\tilde{\tau}_{2}}^{2}\right)\right)\\
   =& -
  \frac{\alpha_{1}}{4\pi}\,M_{1}\,  m_{\tilde{e}_{R}}\, m_{\tilde{\tau}_{R}}\,\delta_{RR}^{13}\, \mu\, \frac{m_{\tau}
\tan\beta}{1+\epsilon_{\tau}\tan\beta}\,  f_{2}\left(M_{1}^{2},\,
      m_{\tilde{e}_{R}}^{2},\, m_{\tilde{\tau}_{1}}^{2},\,
      m_{\tilde{\tau}_{2}}^{2}\right)  .\end{split}
\end{align}
Here and in the following we need the loop functions
\begin{align}\begin{split}
      f_{1}(x,y,z) & = \frac{xy\ln{\frac{x}{y}} + xz\ln{\frac{z}{x}} +
        yz\ln{\frac{y}{z}}}{(x-y)(x-z)(y-z)}\,,\qquad
           f_{2}(x,y,z,w)  = \frac{f_1(x,y,z)-f_1(x,y,w)}{z-w}
      . \label{i-def}\end{split}
\end{align}
For the right diagram in Fig.~\ref{fig:chargedHiggsLFV} we  get the following
contribution:
\begin{align}\label{equ:tauLeRhiggsinobino}\begin{split}
  \Sigma_{\tau_{L}-e_{R}}^{\tilde{H}-\tilde{B}} =&\frac{\alpha_{1}}{4\pi}\,
\frac{m_{\tau}
\tan\beta}{1+\epsilon_{\tau}\tan\beta}\, M_{1}\, \mu\,
\;  m_{\tilde{e}_{R}} m_{\tilde{\tau}_{R}}\delta_{RR}^{13}\cdot \\
& \cdot \left[
\sin^2\theta_\tau f_{2}\left(M_{1}^{2},\,
      \mu^{2},\, m_{\tilde{\tau}_1}^{2},\,
      m_{\tilde{e}_{R}}^{2}\right) + \cos^2\theta_\tau f_{2}\left(M_{1}^{2},\,
      \mu^{2},\, m_{\tilde{\tau}_2}^{2},\,
      m_{\tilde{e}_{R}}^{2}\right)  \right].\end{split}
\end{align}
In order not to get negative slepton masses, $|\delta_{RR}^{13}|$ must for sure
be smaller than 1. Here, we only consider a single flavour-violating mass
insertions. Contributions with double mass insertion,
e.g. $\delta_{LL}^{23}\delta_{RR}^{13*}$ can be relevant for $\mu\to e\gamma$
\cite{Hisano:2009ae}.  In principle, $\Sigma_{\tau_{L}-e_{R}}^{\tilde{\chi}^0} =
\Sigma_{\tau_{L}-e_{R}}^{\tilde{B}} +
\Sigma_{\tau_{L}-e_{R}}^{\tilde{H}-\tilde{B}} $ is sensitive to the RR-element.
However, the relative minus sign is the origin of a possible cancellation in
certain region of the parameter space.  In this approximation the sensitivity to
$\delta_{RR}^{13}$ vanishes if $\mu^2 = m_{\tilde{\tau}_1}^2\cos^2\theta_\tau +
m_{\tilde{\tau}_2}^2\sin^2\theta_\tau $.  In the case with $\theta_\tau = 0$
  the cancellation occurs for $\mu^2 = m_{\tilde{\tau}_L}^2 $. This feature was
already discovered in Ref.~\cite{Paradisi:2006jp}.
In Refs.~\cite{Masiero,Masiero:2008cb} the decoupling limit $M_{\rm SUSY}
  \gg v$ is adopted. In this limit $\tilde\tau_{L,R}$ appear in the loop
  functions instead of $\tilde\tau_{1,2}$ and the $\tilde
  \tau_L$--$\tilde\tau_R$ flip in the bino diagram is incorporated within the
  mass insertion approximation (MIA). This results in a simplified version of
our Eqs.~(\ref{equ:tauLeRbino}) and~(\ref{equ:tauLeRhiggsinobino}); in
\eq{equ:tauLeRhiggsinobino} the square bracket simplifies to
$ f_{2}\left(M_{1}^{2},\, \mu^{2},\, m_{\tilde{\tau}_R}^{2},\,
  m_{\tilde{e}_{R}}^{2}\right)$.\footnote{The relation between the
    notation in Ref.~\cite{Masiero,Masiero:2008cb} and ours is
  $\Sigma_{\tau_L-e_R}= \frac{m_\tau\tan\beta}{1+\epsilon_\tau \tan
    \beta}\Delta_R^{3e} $. E.g.\ the second term in $\Delta_R^{3e}$ corresponds
  to the bino diagram with the following simplification compared to
  \eq{equ:tauLeRbino}: $\sin\theta_\tau\cos\theta_\tau(
  f_1\left(M_{1}^{2},\,m_{\tilde{e}_{R}}^{2},\,m_{\tilde{\tau}_{1}}^{2}\right)-f_1\left(M_{1
    }^{2},\,m_{\tilde{e}_{R}}^{2},\,m_{\tilde{\tau}_{2}}^{2}\right) \approx
  f_1^\prime(M_1^2,m_L^2, m_R^2) $. }  
  Since $\theta_\tau$ vanishes for
  $M_{\rm SUSY} \gg v$, the consideration of large stau mixing requires to go
  beyond the decoupling limit and beyond MIA. Furthermore the stau and bino masses can
  still be smaller than $v$; in fact the interesting region of the parameter
  space probed by NA62 comes with light bino and staus.

We now estimate the maximal allowed LFV effect in $R_K$ including
stau mixing, which depends very much on $\mu\tan\beta$.  At the end of
Sect.~\ref{sect:lfc} we already concluded that effective LFC effects
are typically below the experimental sensitivity. 
In our plots and numerical examples we use the following values for 
the smaller stau mass $m_{\tilde{\tau}_l}$ and the bino mass parameter 
$M_1$: 
\begin{align}
m_{\tilde{\tau}_l}= 120\,\gev, & 
\quad\qquad\qquad M_1= 100\,\gev   
\end{align}
These values are consistent with the experimental lower bounds of
46$\,$GeV for the neutralino masses and $81.9\,$GeV for
$m_{\tilde{\tau}_l}$ \cite{PDG}. The heavier stau mass
$m_{\tilde{\tau}_h}$ is then calculated from the mixing angle
$\theta_\tau$ and $\mu$.  For the off-diagonal element $\Delta
m_{RR}^{13} = m_{\tilde{e}_R} m_{\tilde{\tau}_R}\delta_{RR}^{13}$ we
choose for simplification $m_{\tilde{e}_R} = 200$~GeV, with  
$m_{\tilde{\tau}_R}$ also calculated from $\theta_{\tau}$, $\mu$, and
$m_{\tilde{\tau}_l}$. With this choice the bino diagram increases with
$\mu$, since $m_{\tilde{\tau}_R} $ and $m_{\tilde{\tau}_h} $ increase
too. (Setting instead $\Delta m_{RR}^{13} = m_{\tilde{e}_R}^2
\delta_{RR}^{13}$ would lead to a finite limit for
$\Sigma_{\tau_L-e_R}^{\tilde{B}} $ for $\mu\to \infty$.)  Furthermore, we
choose $\tan\beta = 50$ and $\mu$ to be real and positive.  With this
chosen input parameters we can analyse the dependence of  
$\Delta r^{\mu-e}$ 
on $\theta_\tau$, $\mu$ and $\delta_{RR}^{13}$.  For small
values of $\mu$ the higgsino-bino diagram dominates, but the maximal
value is rather small. 
For large $\mu$ (and other SUSY masses fixed) the higgsino-bino
diagram tends to zero whereas the pure bino-diagram can become
sizeable. Without stau mixing this diagram would not contribute at
all. Thus, in order to get any sizeable effect, especially for large
values of $\mu$, one has to take stau mixing into account.  With this
setup the largest effect in
$\Sigma^{\tilde{B}}_{\tau_{L}-e_{R}}/m_{\tau}$ comes with a relatively
large mixing angle of $\theta_\tau \approx 26^\circ$. In case of the
higgsino-bino diagram, stau mixing is not important, this diagram
  is maximal for small $\mu$, but nevertheless approximately one
order of magnitude smaller than the largest values found for the
pure bino diagram.For $\theta_\tau \approx 26^\circ$ and $\mu>0$
  one has $\tilde{\tau}_l=\tilde{\tau}_1$, which moreover is dominantly
  left-handed.

\begin{figure}
\centering
 \includegraphics[width=7.3cm]{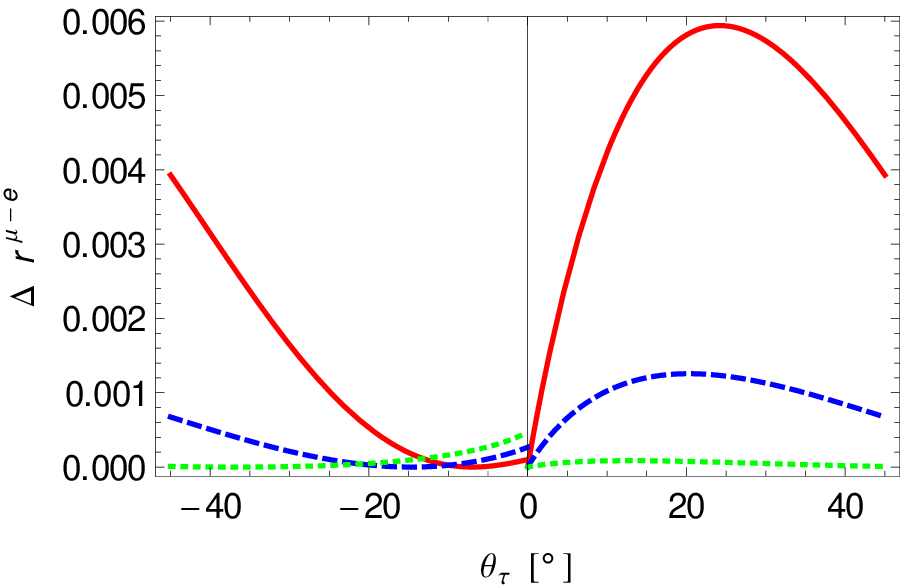}
 \includegraphics[width=7.3cm]{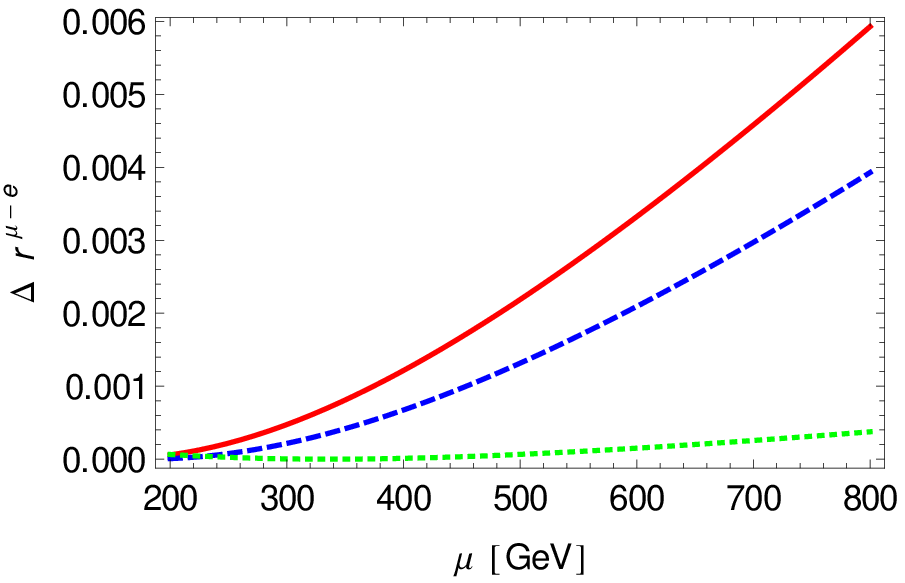}
\caption{$\Delta r^{\mu-e}$ for $\delta_{RR}^{13} = 0.5$, $M_H = 500$~GeV and
$\tan\beta = 50$. Left: As a function of $\theta_\tau$ for different values of $\mu$:
800~GeV (red), 400~GeV (blue
dashed), 200~GeV (green dotted). Right: In dependece of $\mu$ for
different values of $\theta_{\tau}$: $26^\circ$ (red), $45^\circ$ (blue dashed), $
-18^\circ$ (green dotted). }\label{fig:Deltarthetamu}~\\[-2mm]\hrule
\end{figure}

 The maximal possible deviation of $R_K$ from the SM prediction is
 visualized in Fig.~\ref{fig:Deltarthetamu} where $\Delta r^{\mu-e}$
 is plotted for $\delta_{RR}^{13} = 0.5$, $M_H = 500$~GeV and $\tan
   \beta=50$ as a function of $\theta_\tau$ and $\mu$ using typical
 values of $\epsilon_s \tan\beta = 0.3$ and $\epsilon_e \tan\beta
 =\epsilon_\tau\tan\beta =- 0.07$. The discontinuity at $\theta_\tau =
 0$ just comes from the fact that $\tilde{\tau}_1$ and $\tilde{\tau}_2$
 change their roles as heavier and lighter staus.  In order to find
 $\Delta r^{\mu-e}$ for different values of $\delta_{RR}^{13}$,
   $M_H$ and $\tan \beta$ one must rescale those plots using that
   $\Delta r^{\mu-e}$ is quadratic in $\delta_{RR}^{13}$, $\propto
   M_H^{-4}$ and $\propto \tan^6\beta$.  One gets a maximal effect of
 0.6\% for our chosen point of $\mu = 800$~GeV, $M_H = 500$~GeV,
 $\delta_{RR}^{13}= 0.5$, $\theta_\tau = 26^\circ$ and $\tan\beta = 50$,
 which is already in the reach of NA62.  In the range of
 $500\,\rm{GeV}\leq \mu\leq 900\,\rm{GeV}$ a handy approximate
 formula (with an error of 7\%) for the maximal effect (occuring at
   $\theta_\tau = 26^\circ$) can be found
\begin{align}
 \Delta r_{\rm max, LFV}^{\mu-e} \approx \, & 0.006
 \left(\frac{500\,\rm{GeV}}{M_H}\right)^4\left(\frac{\tan\beta}{50}
 \right)^6\left(\frac{\delta_{RR}^{13}}{0.5}\right)^2\left(\frac{\mu}{800\,\rm{GeV}}
 \right)^2. \qquad\qquad \nonumber \\ 
& \qquad\qquad\ \mbox{valid for }
m_{\tilde{\tau}_l}= 120\,\gev, 
\; M_1= 100\,\gev,\;  
m_{\tilde{e}_R} = 200\,\gev .  \label{rapp}
\end{align}
If one varies the lightest stau mass, $ \Delta r^{\mu-e}$ scales
  approximately as $ \lt( 120\,\gev /m_{\tilde{\tau}_l}\rt)^{2}$ in the
  range $ 100\,\rm{GeV}\leq m_{\tilde{\tau}_l}\leq 250\,\rm{GeV}$.  The
  dependence on $M_1$ is roughly linear for $50\,\gev\leq M_1\leq
  100 \gev$ and the prefactor in \eq{rapp} decreases from 0.006 to
  0.0028 if $M_1$ is lowered to 50$\,\gev$. Above 100$\gev$ the $M_1$
  dependence flattens off with a maximum at 200$\gev$, at which the
  prefactor of our approximate formula becomes 0.0078. A full
  investigation of the dependences of $\Delta r^{\mu-e}$ on $M_1$ and
  $m_{\tilde{e}_R} $ requires the use of the exact expression, obtained
  by adding the quantities in
  \eqsand{equ:tauLeRhiggsinobino}{equ:tauLeRbino} to find
  $\Sigma_{\tau_{L}-e_{R}}^{\chi^0}$ and inserting the result into $\Delta
  r^{\mu-e}$ in \eq{rme}.

\begin{figure}
\centering
 \includegraphics[width=7.3cm]{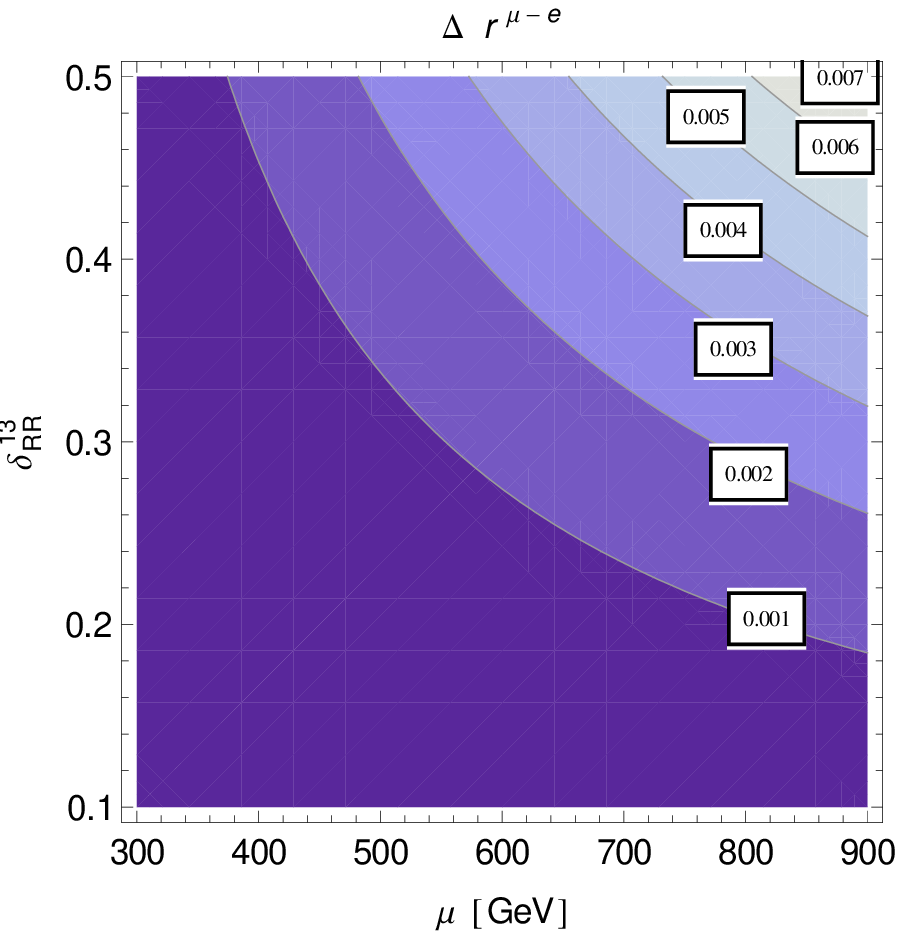}
 \includegraphics[width=7.3cm]{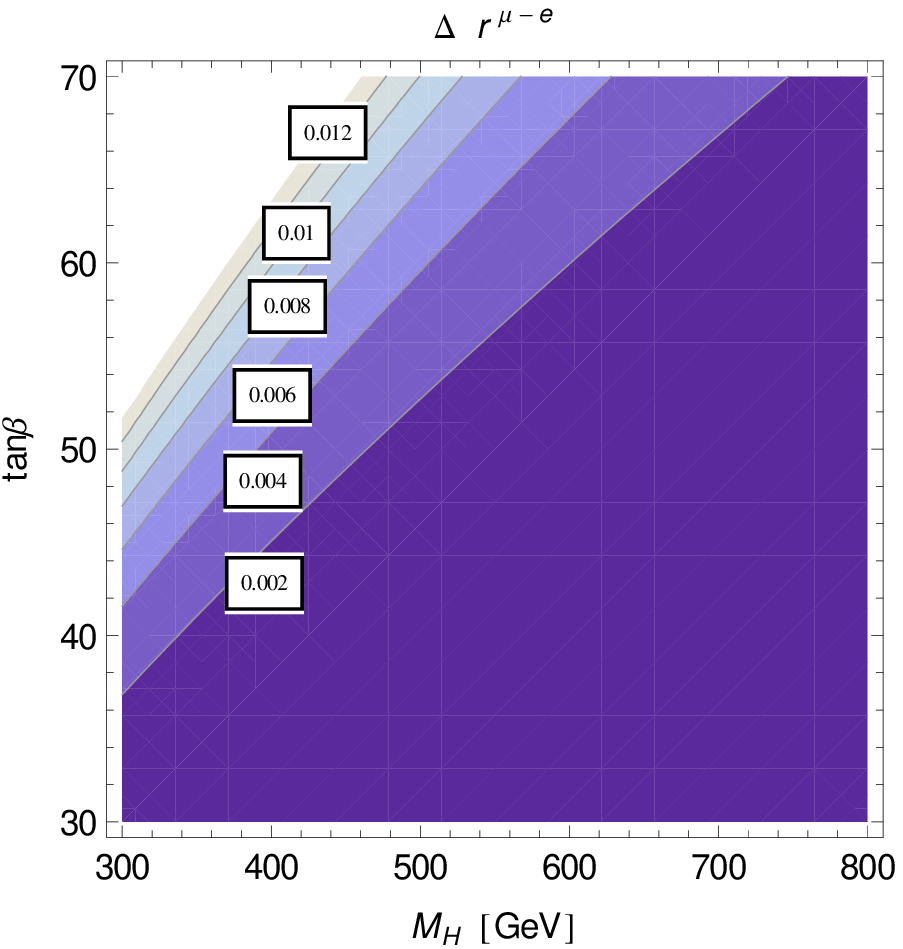}
\caption{$\Delta r^{\mu-e}$ as a function of $\mu$,
  $\delta_{RR}^{13}$, $M_H$ and $\tan\beta$ and stau mixing angle
  $\theta_\tau = 26^\circ$. Left: $M_H = 500$~GeV and $\tan\beta =
  50$. Right: $\mu = 800$~GeV and $\delta_{RR}^{13} =
  0.25$. }\label{fig:deltarmudeltRR13}~\\[-2mm]\hrule
\end{figure}

\begin{figure}
\centering
 \includegraphics[width=7.3cm]{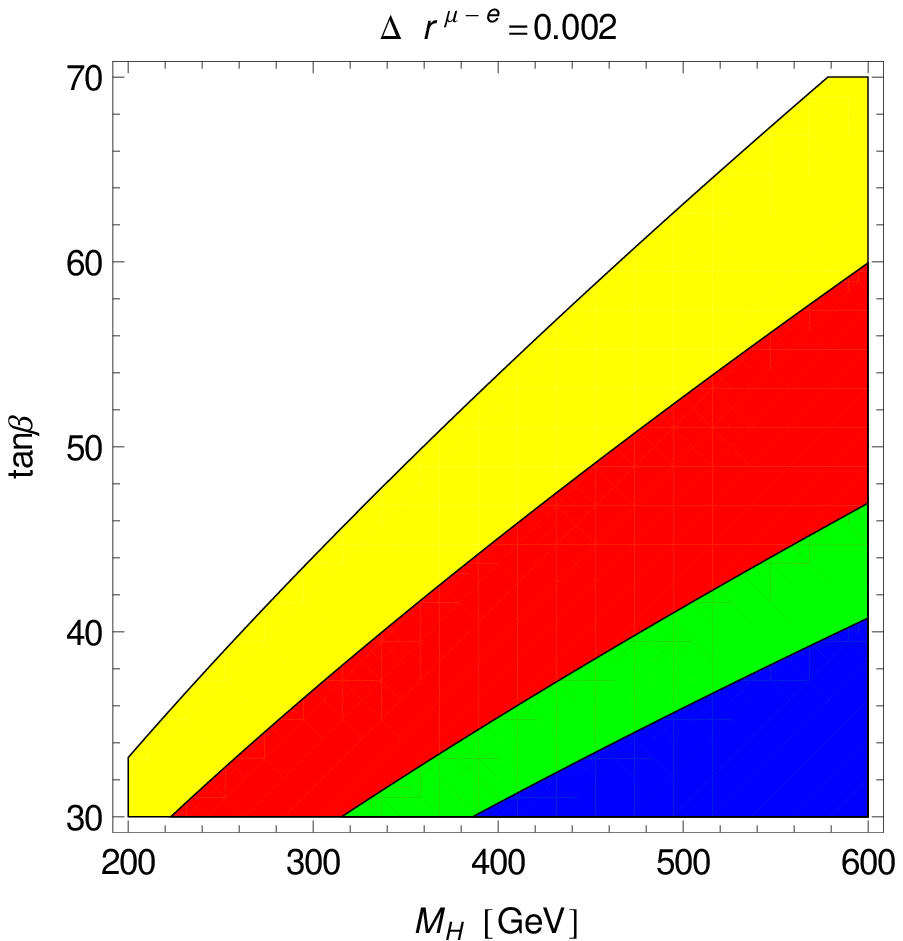}
 \includegraphics[width=7.3cm]{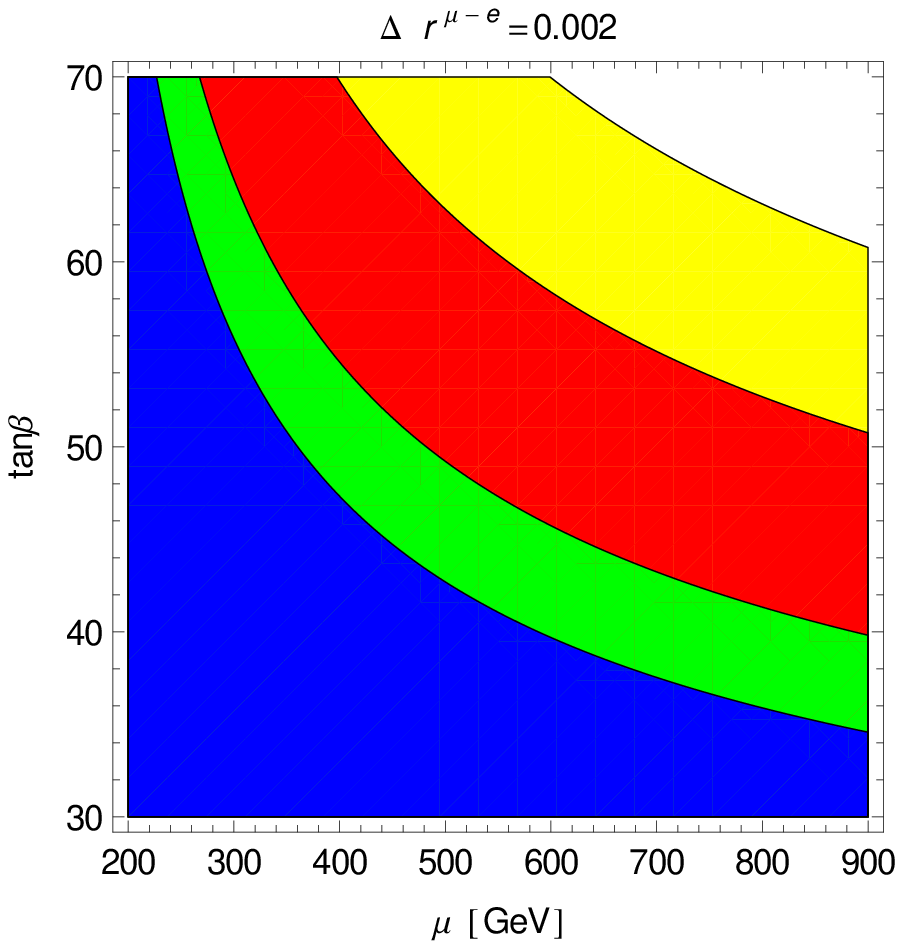}
\caption{For different values of $\delta_{RR}^{13} = 0.15$ (yellow),
  $0.25$ (red), $0.5$ (green), $0.75$ (blue) (from top to bottom) we
  plot the regions in which $\Delta r^{\mu-e}$ is below 
  the future experimental
  sensitivity of 0.002 in the $M_H$--$\tan\beta$ plane with $\mu =
  800$~GeV (left) and in the $\mu$--$\tan\beta$ plane with 
  $M_H = 500$~GeV (right) and
  stau mixing angle $\theta_\tau = 26^\circ$. 
  I.e.\ if $\delta_{RR}^{13} =0.25$, the white and yellow areas
    correspond to $\Delta r^{\mu-e}\geq 0.002$.
}\label{fig:deltartanbetaregion}~\\[-2mm]\hrule
\end{figure}

In \fig{fig:deltarmudeltRR13} we show the dependence of $\Delta
r^{\mu-e}$ on $\mu$, $M_H$, $\tan\beta$ and $\delta_{RR}^{13}$ for
$\theta_\tau = 26^\circ$. It is possible to reach an effect of
$\mathcal{O}(0.5\%)$, whereas for vanishing mixing or small $\mu$ it is
hardly possible to reach the experimental sensitivity. To derive
contraints on $\delta_{RR}^{13}$ Fig.~\ref{fig:deltartanbetaregion}
might be useful. We show regions in the
$\left(M_H,\,\mu,\,\tan\beta,\,\delta_{RR}^{13}\right)$ parameter space
where $\Delta r^{\mu-e}$ reaches the future experimental sensitivity
of 0.2\%. In Ref.~\cite{Girrbach:2009uy} it is pointed out that even large
values of $\tan\beta = 100$ are compatible with the requirement of
pertubative bottom Yukawa coupling.

In order to estimate the contribution of $\tau_L\to e_R$ transition to
$R_K$ in Ref.~\cite{Masiero}, the authors used $\Delta_R^{31} = 5\cdot
10^{-4} $ which is related to our notation by $\Sigma_{\tau_{L}-e_{R}} =
\frac{m_\tau\tan\beta}{1 + \epsilon_\tau \tan\beta}\Delta_R^{31} $ or
$\left|\Sigma^{\tilde{\chi}^0}_{\tau_{L}-e_{R}}\right|/m_{\tau} = 0.025
$.  However, such large values correspond to quite specific points
in the MSSM parameter space, especially extremely large $\mu$.
Thus, in order not to overestimate the effect and to avoid trouble
with too small slepton masses one should rather take at most
$\left|\Sigma^{\tilde{\chi}^0}_{\tau_{L}-e_{R}}\right|/m_{\tau}\approx
0.01$ (meaning $\Delta_{R,\rm max}^{31}= 2\cdot 10^{-4}$).  Taking this
maximal value and further $\tan\beta=50$ and a charged-Higgs mass
of $M_H = 500~$GeV we end up with $\Delta r^{\mu-e}_\text{max,LFV}
\approx 0.007$, which is within the experimental sensitivity of the NA62
experiment.

\begin{figure}
\centering
 \includegraphics[width = 7cm]{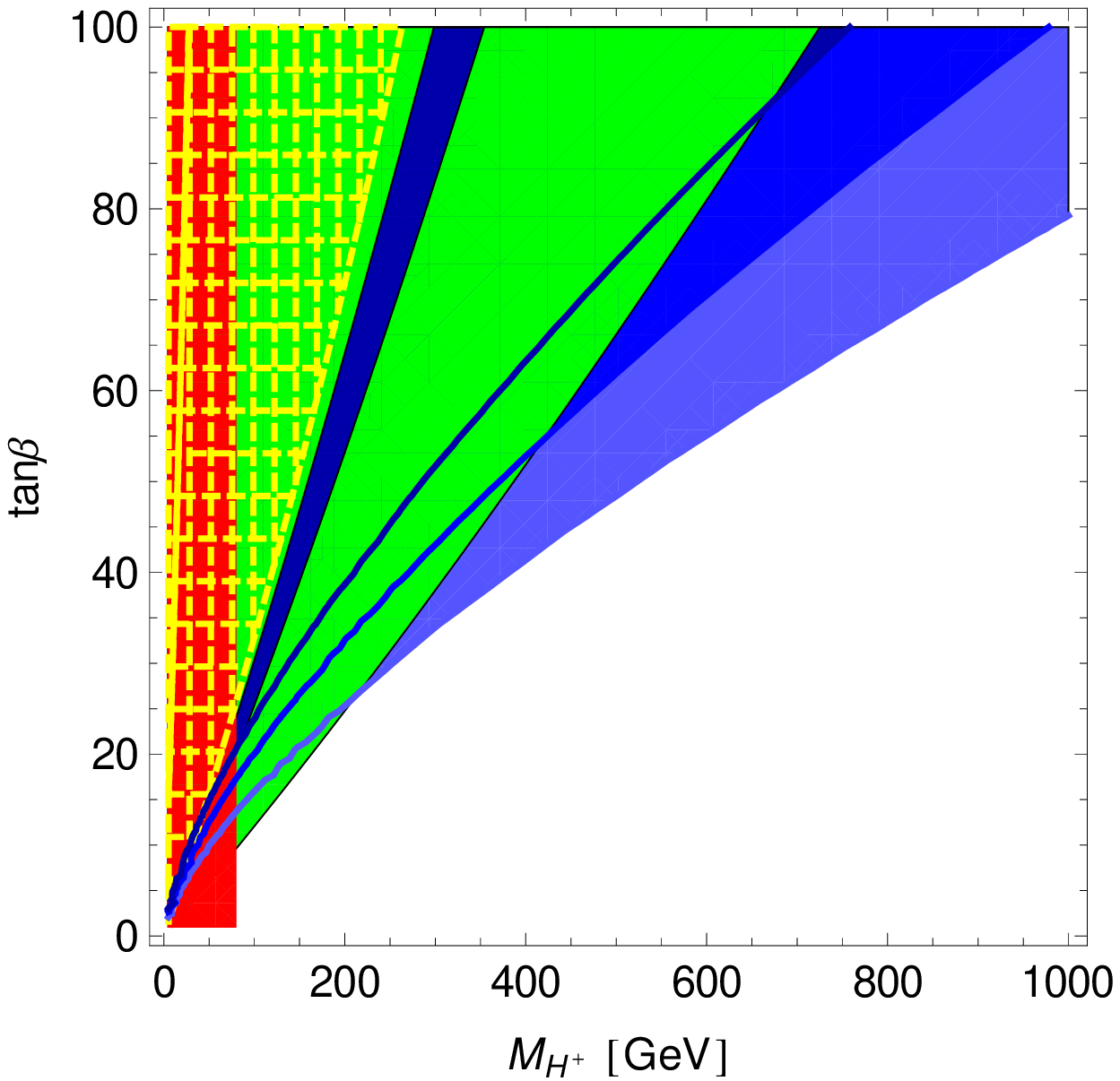}
 \includegraphics[width = 7cm]{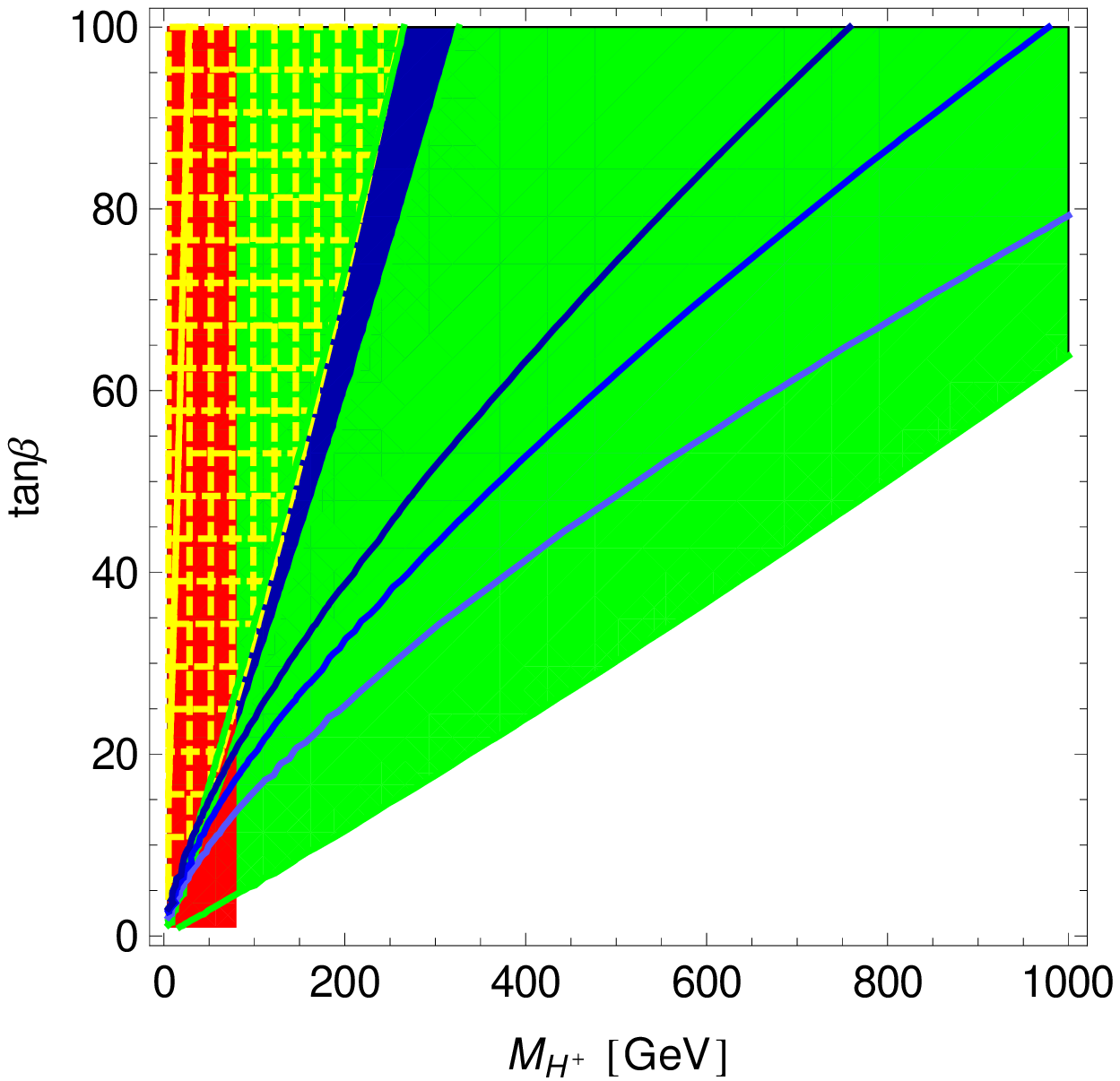}
 \caption{Regions with $\Delta r^{\mu-e} \geq 0.5\%$ for
   $\delta_{RR}^{13} =$ 0.15 (darkblue), 0.25 (blue and darkblue), 0.5
   (lightblue, blue, and darkblue). Overlaid in 
   red: exluded by LEP $H^+$ searches; yellow
   dashed: $3\sigma$ exclusion limit from $R_{\mu23}$; 
   green: $3\sigma$ exclusion limit from
   $B\to\tau\nu$ (left: using $|V_{ub}| = \left(5.04\pm 0.64\right)\cdot
   10^{-3}$; 
   right: using $|V_{ub}| = \left(3.41\pm
   0.15\right)\cdot 10^{-3}$).
   }\label{fig:final}~\\[-2mm]\hrule
\end{figure}

In Ref.~\cite{Goudzovski:2011talk} the 95\%CL exclusion region for $R_K$
for three different values of $\Delta_R^{31}$ ($1\cdot 10^{-3},\,5\cdot
10^{-4},\, 1\cdot 10^{-4}$) is shown in the $(M_{H^\pm},\tan\beta)$
region and compared with the constraints from $B\to\tau\nu$,
$B\to X_s\gamma$, $R_{\mu23}=\Gamma(K\to \mu\nu)/\Gamma(K\to \pi^0 \mu\nu)$
\cite{Antonelli:2010yf} and direct $H^\pm$ searches. According to the
discussion in the preceding paragraph we prefer to take
$\Delta_R^{31} = 2\cdot 10^{-4}$ as the maximal value, so that our
  excluded $(M_{H^\pm},\tan\beta)$ region is smaller than the one in
  Ref.~\cite{Goudzovski:2011talk}.  In the following we use
\cite{Uli:2011talk,Nierste:2011na,Barlow:2011fu,Nierste:2008qe}
\begin{align}
& \mathcal B(B\to\tau\nu)^\text{SM} = 1.13\cdot 10^{-4}
  \left(\tfrac{|V_{ub}|}{4\cdot
    10^{-3}}\right)^2\left(\tfrac{f_B}{200~\text{MeV}}\right)^2\,,\\ & \mathcal B(B\to
  \tau\nu)^\text{exp} = \left(1.64\pm 0.34\right)\cdot
  10^{-4}\,,\\ &  \mathcal B(B\to \tau\nu)^\text{SUSY} =
  \left[1-\left(\frac{m_B}{m_{H^+}}\right)^2\frac{\tan^2\beta}{(1+\epsilon_0
      \tan\beta)(1+\epsilon_\tau \tan\beta)}\right]^2  \mathcal B(B\to
  \tau\nu)^\text{SM}
\end{align}
with $\epsilon_0 \approx \epsilon_s\approx \tfrac{1}{16\pi^2}$
$\epsilon_\tau\tan\beta = -0.07$and $\mu=800\,\gev$ as above.  In
Fig.~\ref{fig:final} we plot the region in the in the
  $M_{H^\pm}$--$\,\tan\beta$ plane satisfying $\Delta r^{\mu-e} \leq
0.5\%$ for three different values of $\delta_{RR}^{13}$ (0.15, 0.25,
0.5). Overlaid are the constraints from $B\to\tau\nu$,
$R_{\mu23}= \Gamma(K\to \mu\nu)/\Gamma(K\to \pi^0 \mu\nu)$ and direct $H^\pm$
searches. The prediction of $\mathcal B(B\to\tau\nu)$ within the SM and the
  MSSM requires the knowledge of $|V_{ub}|$.  Determinations of
  $|V_{ub}|$ from different quantities result in substantially different
  numerical predictions. For discussions of this ``$V_{ub}$-puzzle'' see
  Refs.~\cite{Uli:2011talk,Nierste:2008qe,Lenz:2010gu}.  For our
  analysis we consider two extreme scenarios: First, in the left plot
of Fig.~\ref{fig:final} $|V_{ub}|$ is determined such that the SM
prediction of $\mathcal B(B\to\tau\nu)$ is equal to the experimental value.
Using $f_B = \left(191\pm13\right)~\text{MeV}$ one gets $|V_{ub}| =
\left(5.04\pm 0.64\right)\cdot 10^{-3}$ \cite{Uli:2011talk}. In the plot
we set $\mathcal B(B\to\tau\nu)^\text{SM} =\mathcal B(B\to\tau\nu)^\text{exp} $ and
use the experimental $3\sigma$-region. Second, in the right plot of
Fig.~\ref{fig:final} $V_{ub}$ is fixed to the best-fit value of a
  global fit to the unitarity triangle \cite{Lenz:2010gu}. An essential
  assumption of the second scenario is the absence of new physics in the
  CP asymmetry $A_{\rm CP}^{\rm mix}(B\to J/\psi K_S)$, from which the
  angle $\beta$ of the unitarity triangle is determined: Then
  $|V_{ub}|\propto |V_{cb}|\tfrac{\sin\beta}{\sin\alpha}$ leads to
$|V_{ub}|=\left(3.41\pm 0.15\right)\cdot 10^{-3} $
\cite{Uli:2011talk}. In this case the SM central value
$\mathcal B(B\to\tau\nu)^\text{SM} = 0.75\cdot 10^{-4}$ is much lower than the
experimental value and the SUSY contribution makes it even
smaller. Therefore now $\mathcal B(B\to\tau\nu)$ is much more constraining
  than in the scenario of the left plot.
Using the $2\sigma$ region of the experimental
value one would exclude the whole $(M_{H^\pm},\tan\beta)$ region 
(except for a very narrow strip with $\tan\beta\approx 0.3 M_{H^+}$); the
constraint from the $3\sigma$ region is shown in the right plot
of \fig{fig:final}.  In the future also $\mathcal B(B\to D \tau \nu)$
\cite{Nierste:2008qe,Kamenik:2008tj,Tanaka:2010se} and $\mathcal B(B\to \pi \tau
\nu)$ \cite{Khodjamirian:2011ub} will probe charged-Higgs effects and
will eventually shed light on the situation.

We finally mention two related studies: A prospective error of
0.12\% of the NA62 experiment at CERN is used in
Ref.~\cite{Ellis:2008st}.  The parameter scan in this paper
respects $\left|\delta_{LL}^{13}\delta_{RR}^{13}\right|\leq 0.01$, in
agreement with our result. In Ref.~\cite{Filipuzzi:2009xr} it is
pointed out, using a general effective theory approach, that in models
with Minimal Lepton Flavour Violation (MLFV and MFV-GUT) the effects are
too small to be observed.

\section{Conclusions\label{sect:c}}
The NA62 experiment has the potential to discover new sources of
  lepton flavour violation by testing lepton flavour universality
  through a precision measurement of $R_K=\Gamma\left(K\to
  e\nu\right)/\Gamma\left(K\to\mu\nu\right)$ \cite{Masiero}. This kind
  of new physics dominantly affects the decay rate $\Gamma(K\to e
  \nu)$. A lepton-flavour conserving (LFC) mechanism changing
  $\Gamma(K\to e \nu_e)$ may suppress or enhance $R_K$, while new
  lepton-flavour violating (LFV) decay modes such as $\Gamma(K\to e
  \nu_\tau)$ can only enhance $R_K$ over its SM value. In this paper we
  have studied $\Delta r^{\mu-e}\equiv R_K/R_{K}^{\rm SM}-1$ in the MSSM, 
  extending the analyses of Refs.~\cite{Masiero,Masiero:2008cb}.
 
The LFC contribution to $\Delta r^{\mu-e}$ is driven by the
  parameter combination $\delta_{LL}^{13}\delta_{RR}^{13}$. In
  Ref.~\cite{Girrbach:2009uy} it has been found that upper bounds on
  $|\delta_{LL}^{13}\delta_{RR}^{13}|$ can be derived from naturalness
  considerations of the electron mass and from the precise measurement
  of the anomalous magnetic moment of the electron. (Coincidentally,
  these two quantities give very similar constraints.) In
  Sect.~\ref{sect:lfc} we have found that these bounds imply $|\Delta
  r^{\mu-e}_\text{LFC}|\lesssim 0.005$ and thereby challenge the large
  values for $|\Delta r^{\mu-e}_\text{LFC}|$ considered in
  Ref.~\cite{Masiero}. At the same time our result is fully compatible
  with the range for $\Delta r^{\mu-e}_\text{LFC}$ advocated in
  Ref.~\cite{Masiero:2008cb}. The naturalness bound extends beyond the
  MSSM to a larger class of models, namely those with the tree-level
  Higgs sector of a 2HDM of type II.

  The LFV contribution to $\Delta r^{\mu-e}$ can be larger, because a non-zero
  parameter $\delta_{RR}^{13}$ suffices to open the decay channel $K\to e
  \nu_\tau$ and $\delta_{RR}^{13}$ is only poorly constrained from other
  processes. We have calculated $\Delta r^{\mu-e}_{\text{LFV}}$ in
  Sect.~\ref{sec:LFVself-energy} and found that the proper inclusion of
  $\tilde{\tau}_L$--$\tilde{\tau}_R$ mixing is essential. The analytical
  expressions in Refs.~\cite{Masiero,Masiero:2008cb} include the
  $\tilde\tau_L-\tilde\tau_R$ flip using the mass insertion approximation
  instead of the exact diagonalisation of the stau mass matrix.  The
  interesting region of parameter space probed by NA62 corresponds to large
  values of $\mu$ and a sizable stau mixing angle $\theta_\tau$ and in this
  region the left (bino) diagram in \fig{fig:chargedHiggsLFV} is
  dominant. The formulae derived by us are also valid beyond the decoupling
    limit $M_{\rm SUSY}\to \infty$, in which $\theta_\tau$ vanishes. In order
  to facilitate the combination of future NA62 results with limits or
  measurements from high-$p_T$ experiments, we have expressed $\Delta
  r^{\mu-e}_{\text{LFV}}$ in terms of the mass $m_{\tilde{\tau}_l}$ of the
  lightest stau eigenstate and the mixing angle $\theta_\tau$.  For example, for
  $\tan\beta = 50$, $\mu = 800~$GeV, $\delta_{RR}^{13} = 0.5$, a charged-Higgs
  mass of $M_H = 500~$GeV, $m_{\tilde{\tau}_l}= 120\,\gev$, a bino mass of $M_1=
  100\,\gev$ and a right-handed selectron mass of $m_{\tilde{e}_R} = 200\,\gev$
  we find a maximal value of $\Delta r^{\mu-e}_{\text{LFV}}=0.006 $
  corresponding to $\theta_\tau = 26^\circ$. In \eq{rapp} we have derived an
  easy-to-use formula expressing $\Delta r^{\mu-e}_{\text{LFV}}$ in terms of the
  relevant MSSM parameters. Finally we have plotted the regions of the MSSM
  parameter space probed by $R_K$ and briefly compared the result with the
  constraint from other observables such as $\mathcal B(B\to \tau \nu)$.
  
\section*{Acknowledgements}
The presented work is supported by BMBF Grant No.05H09VKF.
J.G.\ acknowledges the financial support by \emph{Studienstiftung des
  deutschen Volkes} and the DFG cluster of excellence
\lq\lq Origin and Structure of the Universe''.
U.N.\ is grateful for the hospitality of the \emph{Institute for
    Advanced Study}\ of Technische Universit\"at M\"unchen, where this
  paper has been completed.

\bibliography{LitKaonPaper}
\bibliographystyle{JHEP}
\end{document}